\documentclass{ws-procs9x6}

\begin{document}

\title{
  \vspace*{-0.5cm}
  \hfill{\normalsize\vbox{%
    \hbox{\rm\small DPNU-04-03}
  }}\\
  \vspace{0.2cm}
 Pion Velocity \\ near the Chiral Phase Transition~\footnote{%
 \uppercase{T}alk given at \uppercase{KIAS-APCTP} 
 \uppercase{I}nternational \uppercase{S}ymposium on 
 \uppercase{A}stro-\uppercase{H}adron \uppercase{P}hysics, 
 \uppercase{N}ovember 10-14, 2003, 
 \uppercase{KIAS}, \uppercase{S}eoul, \uppercase{K}orea.
 \uppercase{T}his talk is based on the works done in 
 \uppercase{R}efs.~1 and 2.
}
}

\author{Chihiro Sasaki}

\address{Department of Physics, Nagoya University,
Nagoya, 464-8602, JAPAN
}

\maketitle

\abstracts{
We study the pion velocity near the critical temperature $T_c$ of chiral
symmetry restoration in QCD. 
Using the hidden local symmetry (HLS) model as
the effective field theory, where the chiral symmetry restoration
is realized as the vector manifestation (VM), we show that the
pion velocity for $T \to T_c$ receives neither quantum
nor (thermal) hadronic corrections at the critical temperature
even when we start from the bare theory with Lorentz symmetry
breaking. We show that this is related to a new fixed point
structure originated in the VM.
Further we match at a matching scale
the axial-vector current correlator in the HLS with
the one in the operator product expansion for QCD, and present the
matching condition to determine the bare pion velocity. We find
that the pion velocity is
close to the speed of light,
$v_\pi (T) = 0.83 - 0.99$.
}

\section{Introduction}

Chiral symmetry in QCD is expected to be restored under some
extreme conditions such as large number of flavor $N_f$ and high
temperature and/or density. In hadronic sector, the chiral
symmetry restoration is described by various effective field
theories (EFTs)
based on the chiral symmetry ~\cite{review:chiral}. 

By using the hidden local symmetry (HLS)
model~\cite{Bando} as an EFT
and performing the Wilsonian matching which is one of the methods
that determine the bare theory from the underlying QCD~\cite{HY:WM}, 
the vector manifestation (VM) in hot
or dense matter was formulated in Refs.~\refcite{HS:VM,HKR:VM}. 
In the VM,
the massless vector meson becomes the chiral partner of pion at
the critical point~\cite{HY:VM}~\footnote{
 As studied in Ref.~\refcite{HY:PR} in detail,
 the VM is defined only as a limit with bare parameters 
 approaching the VM fixed point from the broken phase.
}.
There, the {\it intrinsic temperature or
density dependences of the parameters} of the HLS Lagrangian,
which are obtained by integrating out the high energy modes
(i.e., the quarks and gluons above the matching scale) in hot
and/or dense matter, play
the essential roles to realize the chiral symmetry restoration
consistently.
That the
vector meson mass vanishes at the critical temperature/density
supports the in-medium scaling of the vector meson proposed by
Brown and Rho, Brown-Rho scaling~\cite{BRscaling}, and has
qualitatively important influences on the properties of hadrons in
medium.

In the analysis done in
Ref.~\refcite{HKRS:SUS}, it was shown that the effect of Lorentz
symmetry breaking to the bare parameters caused by the intrinsic
temperature dependence through the Wilsonian matching are
small~\cite{HKRS:SUS,HS:VVD}. 
Starting from the bare Lagrangian with Lorentz invariance, 
it was presented that the pion velocity
approaches the speed of light at the critical temperature~\cite{HKRS:SUS}, 
although in low temperature region $(T \ll
T_c)$ the pion velocity deviates from the speed of light due to
hadronic corrections~\cite{HS:VVD}.

However there do exist the Lorentz non-invariant effects 
in bare EFT anyway
due to the intrinsic temperaure and/or density effects.
Further the Lorentz non-invariance might be enhanced through the 
renormalization group equations (RGEs),
even if effects of Lorentz symmetry breaking at the bare level are small.
Thus it is important to investigate how Lorentz non-invariance 
at bare level influences physical quantities.  

In this talk, 
we pick up the pion velocity at the critical temperature
and study the quantum and hadronic thermal effects
based on the VM.
The pion velocity is one of the important quantities since it controls
the pion propagation in medium through a dispersion relation.
We show the non-renormalization property on the pion velocity $v_\pi$,
which is protected by the VM, and estimate the value of $v_\pi$
near the critical temperature.

\section{Model Based on the Hidden Local Symmetry}

In this section, we show the HLS Lagrangian at leading order
including the effects of Lorentz non-invariance.

The HLS model is based on
the $G_{\rm{global}} \times H_{\rm{local}}$ symmetry,
where $G=SU(N_f)_L \times SU(N_f)_R$ is the chiral symmetry
and $H=SU(N_f)_V$ is the HLS.
The basic quantities are
the HLS gauge boson $V_\mu$ and two matrix valued
variables $\xi_L(x)$ and $\xi_R(x)$
which transform as
$\xi_{L,R}(x) \to \xi^{\prime}_{L,R}(x)
  =h(x)\xi_{L,R}(x)g^{\dagger}_{L,R}$,
where $h(x)\in H_{\rm{local}}\ \mbox{and}\ g_{L,R}\in
[SU (N_f)_{\rm L,R}]_{\rm{global}}$.
These variables are parameterized as
\footnote{
 The wave function renormalization constant of the pion field
 is given by the temporal component of the pion decay constant
 ~\cite{sasaki,Meissner:2001gz}.
 Thus we normalize $\pi$ and $\sigma$ by $F_\pi^t$ and $F_\sigma^t$
 respectively.
}
%\vspace*{-0.2cm}
$ \xi_{L,R}(x)=e^{i\sigma (x)/{F_\sigma^t}}
     e^{\mp i\pi (x)/{F_\pi^t}}$,
%\vspace*{-0.2cm}
where $\pi = \pi^a T_a$ denotes the pseudoscalar Nambu-Goldstone
bosons associated with the spontaneous symmetry breaking of
$G_{\rm{global}}$ chiral symmetry, and $\sigma = \sigma^a T_a$
denotes the Nambu-Goldstone bosons associated with the spontaneous
breaking of $H_{\rm{local}}$. This $\sigma$ is absorbed into the
HLS gauge boson through the Higgs mechanism, and then the vector
meson acquires its mass. $F_\pi^t$ and $F_\sigma^t$ denote
the temporal components of the decay constant of
$\pi$ and $\sigma$, respectively.
The covariant derivative of $\xi_{L}$ is given
by
%\vspace*{-0.2cm}
\begin{equation}
 D_\mu \xi_L = \partial_\mu\xi_L - iV_\mu \xi_L + i\xi_L{\mathcal{L}}_\mu,
\end{equation}
%\vspace*{-0.2cm}
and the covariant derivative of $\xi_R$ is obtained
by the replacement of ${\mathcal L}_\mu$ with ${\mathcal R}_\mu$
in the above where
$V_\mu$ is the gauge field of $H_{\rm{local}}$, and
${\mathcal{L}}_\mu$ and ${\mathcal{R}}_\mu$ are the external
gauge fields introduced by gauging $G_{\rm{global}}$ symmetry.
In terms of ${\mathcal L}_\mu$ and ${\mathcal R}_\mu$,
we define the external axial-vector and vector fields as
${\mathcal A}_\mu = ( {\mathcal R}_\mu - {\mathcal L}_\mu )/2$ and
${\mathcal V}_\mu = ( {\mathcal R}_\mu + {\mathcal L}_\mu )/2$.

In the HLS model it is possible to perform the derivative
expansion systematically~\cite{Georgi,Tanabashi:1993sr,HY:PR}. 
In the chiral perturbation theory (ChPT) with  HLS, 
the vector meson mass is to be considered as small compared with 
the chiral symmetry breaking scale $\Lambda_\chi$, 
by assigning ${\mathcal O}(p)$ to the HLS gauge coupling, 
$g \sim {\mathcal O}(p)$~\cite{Georgi,Tanabashi:1993sr}. 
(For details of the ChPT with HLS, see Ref.~\refcite{HY:PR}.) 
The leading order Lagrangian with
Lorentz non-invariance can be written as~\cite{HKR:VM}
%\vspace*{-0.1cm}
\begin{eqnarray}
&&
{\mathcal L}
=
\biggl[
  (F_{\pi}^t)^2 u_\mu u_\nu
  {}+ F_{\pi}^t F_{\pi}^s
    \left( g_{\mu\nu} - u_\mu u_\nu \right)
\biggr]
\mbox{tr}
\left[
  \hat{\alpha}_\perp^\mu \hat{\alpha}_\perp^\nu
\right] \nonumber\\
&&\qquad{}+
\biggl[
  (F_{\sigma}^t)^2 u_\mu u_\nu
  {}+  F_{\sigma}^t F_{\sigma}^s
    \left( g_{\mu\nu} - u_\mu u_\nu \right)
\biggr]
\mbox{tr}
\left[
  \hat{\alpha}_\parallel^\mu \hat{\alpha}_\parallel^\nu
\right]
\nonumber\\
&&\qquad
{} +
\Biggl[
  - \frac{1}{ g_{L}^2 } \, u_\mu u_\alpha g_{\nu\beta}
  {}- \frac{1}{ 2 g_{T}^2 }
  \left(
    g_{\mu\alpha} g_{\nu\beta}
   - 2 u_\mu u_\alpha g_{\nu\beta}
  \right)
\Biggr]
\mbox{tr}
\left[ V^{\mu\nu} V^{\alpha\beta} \right]
\ ,
\label{Lag}
\end{eqnarray}
%\vspace*{-0.1cm}
where
%\vspace*{-0.1cm}
\begin{equation}
 \hat{\alpha}_{\perp,\parallel }^{\mu}
 = \frac{1}{2i}\bigl[ D^\mu\xi_R \cdot \xi_R^{\dagger}
                 {}\mp  D^\mu\xi_L \cdot \xi_L^{\dagger}
                   \bigr].
\end{equation}
%\vspace*{-0.1cm}
Here $F_{\pi}^s$ denote the spatial pion decay constant and similarly
$F_{\sigma}^s$ for the $\sigma$. The rest frame
of the medium is specified by $u^\mu = (1,\vec{0})$ and
$V_{\mu\nu}$ is the field strength of $V_\mu$. $g_{L}$ and $g_{T}$
correspond in medium to the HLS gauge coupling $g$. The parametric
$\pi$ and $\sigma$ velocities are defined by~\cite{Pisarski:1996mt}
%\vspace*{-0.1cm}
\begin{equation}
 V_\pi^2 = {F_\pi^s}/{F_\pi^t}, \qquad
 V_\sigma^2 = {F_\sigma^s}/{F_\sigma^t}.
\end{equation}

\section{Vector Manifestation Conditions}

In this section, 
we start from the HLS Lagrangian with Lorentz
non-invariance, and requiring that the axial-vector current correlator be
equal to the vector current correlator at the critical point, we
present the conditions satisfied at the critical point.

Concept of the matching in the Wilsonian sense is based on the following
assumptions:
The bare Lagrangian of the effective field theory (EFT) 
${\mathcal L}_{\rm bare}$ is defined at a suitable matching scale $\Lambda$.
Generating functional derived from ${\mathcal L}_{\rm bare}$ leads to
the same Green's function as that derived from the generating functional
of QCD at $\Lambda$.
In other words, the bare parameters are obtained after integrating out
the high energy modes, i.e., the quarks and gluons above $\Lambda$.
When we integrate out the high energy modes in hot matter,
the bare parameters have a certain 
temperature dependence, {\it intrinsic temperature dependence}, 
converted from QCD to the EFT.
The intrinsic temperature dependence is nothing but the signature
that hadrons have an internal structure constructed from 
quarks and gluons.
In the following, we describe the chiral symmetry restoration
based on the point of view that {\it the bare HLS theory is
defined from the underlying QCD.}
We note that the Lorentz non-invariance appears in bare HLS theory
as a result of including the intrinsic temperature dependences.
Once the temperature dependence of the bare parameters is determined 
through the matching with QCD mentioned above,
from the RGEs the parameters 
appearing in the hadronic corrections pick up the intrinsic 
thermal effects.

Now we consider the matching near the critical temperature.
The axial-vector and vector current correlators at bare level
are constructed in terms of bare parameters and
are expanded in terms of the longitudinal and transverse
projection operators $P_{L,T}^{\mu\nu}$:
$G_{A,V}^{\mu\nu} = P_L^{\mu\nu}G_{A,V}^L + P_T^{\mu\nu}G_{A,V}^T$.
At the chiral phase transition point, the axial-vector and vector
current correlators must agree with each other,
i.e., chiral symmetry restoration is characterized by 
the following conditions:
$G_{A{\rm(HLS)}}^L - G_{V{\rm(HLS)}}^L \to 0$ and 
$G_{A{\rm(HLS)}}^T - G_{V{\rm(HLS)}}^T \to 0$ for $T \to T_c$.
In Ref.~\refcite{HKR:VM}, it was shown that
they are satisfied for
any values of $p_0$ and $\bar{p}$ around the matching scale
only if the following conditions are met:
$(g_{L,{\rm bare}}, g_{T,{\rm bare}}, a_{\rm bare}^t, a_{\rm bare}^s)
\to (0,0,1,1)$ for $T \to T_c$.
This implies that at bare level the longitudinal mode of the vector 
meson becomes the real NG boson and couples to the vector current correlator, 
while the transverse mode decouples.

In Ref.~\refcite{sasaki}, we have shown that
$(g_L, a^t, a^s) = (0,1,1)$ is a fixed point of the RGEs and 
satisfied in any energy scale. 
Thus the VM condition is given by
\begin{eqnarray}
 (g_L, a^t, a^s) \to (0,1,1) \quad \mbox{for}\quad T \to T_c.
\label{evm}
\end{eqnarray}
The vector meson mass is never generated at the critical temperature
since the quantum correction to $M_\rho^2$ is proportional to $g_L^2$.
Because of $g_L \to 0$,
the transverse vector meson at the critical point, in any energy scale, 
decouples from the vector current correlator.
The VM condition for $a^t$ and $a^s$ leads to the equality between
the $\pi$ and $\sigma$ (i.e., longitudinal vector meson) velocities:
\begin{eqnarray}
 \bigl( V_\pi / V_\sigma \bigr)^4
 = \bigl( F_\pi^s F_\sigma^t / F_\sigma^s F_\pi^t \bigr)^2
 = a^t / a^s 
 \stackrel{T \to T_c}{\to} 1.
\label{vp=vs}
\end{eqnarray}
This is easily understood from a point of view of the VM
since the longitudinal vector meson becomes the chiral partner
of pion.
We note that this condition
$V_\sigma = V_\pi$ holds independently of the value of the bare
pion velocity which is to be determined through the Wilsonian matching.

\section{Non-renormalization Property on the Pion Velocity}

As we have seen in the previous section,
the dropping mass of vector meson can be realized as the VM
which is formulated by using the HLS theory.
Then what is the prediction of the VM?
Recently it was proven that the non-renormalization property
on the pion velocity which is protected by the VM~\cite{sasaki}.
In the following, we show that this can be understood based on the idea
of chiral partners.

\begin{figure}
 \begin{center}
  \includegraphics[width = 7cm]{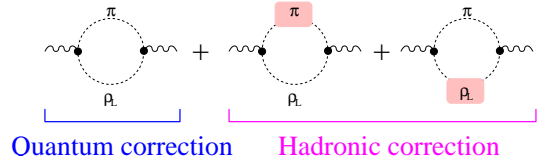}
 \end{center}
 \caption{Diagrams for contribution to the pion velocity.
 Center (right) diagram denotes the interaction between the pion 
 (vector meson) and heat bath.}
 \label{pi-rho}
\end{figure}

Before going to the critical temperature $T_c$, 
let us consider the situation away from $T_c$.
Starting from the bare pion velocity 
$V_{\pi,{\rm bare}}^2 = F_{\pi,{\rm bare}}^s / F_{\pi,{\rm bare}}^t$ 
and including quantum and hadronic corrections into the parameters
through the diagrams shown in Fig.~\ref{pi-rho}.
Away from $T_c$, there exists the hadronic thermal correction to 
the pion velocity~\cite{HS:VVD}:
\begin{equation}
 v_\pi^2 (T) \simeq
 V_\pi^2 - N_f \frac{2\pi^2}{15}\frac{T^4}{(F_\pi^t)^2 M_\rho^2}
 \qquad \mbox{for} \quad T < T_c,
\label{low T}
\end{equation}
where the contribution of the massive 
$\sigma$ (i.e., the longitudinal mode of massive vector meson) is 
suppressed owing to the Boltzmann factor $\exp [-M_\rho / T]$, 
and then only the pion loop contributes to the pion velocity.

On the other hand,
when we approach the critical temperature,
the vector meson mass goes to zero due to the VM.
Thus $\exp [-M_\rho / T]$ is no longer the suppression factor.
As a result, the hadronic correction in the pion velocity is absent
due to the exact cancellation between the contribution of pion and 
that of its chiral partner $\sigma$.
Similarly the quantum correction generated from the pion loop 
is exactly cancelled by that from the $\sigma$ loop.
Accordingly we conclude
\begin{equation}
 v_\pi(T) = V_{\pi,{\rm bare}}(T)
 \qquad \mbox{for}\quad T \to T_c,
\label{phys=bare}
\end{equation}
i.e., {\it the pion velocity in the limit $T \to T_c$
receives neither hadronic nor quantum corrections due
to the protection by the VM.}
This implies that $(g_L,a^t,a^s,V_\pi) = (0,1,1,\mbox{any})$ forms
a fixed line for four RGEs of $g_L, a^t, a^s$ and $V_\pi$.
When one point on this fixed line is selected through the matching
procedure as done in Ref.~\refcite{HKRS:pv},
namely the value of $V_{\pi,{\rm bare}}$ is fixed,
the present result implies that the point does not move in a subspace
of the parameters. 
Approaching the restoration point of chiral
symmetry, the physical pion velocity itself flows into the fixed
point.

\section{Matching Conditions on the Bare Pion Velocity}

One possible way to determine the bare parameters is the
Wilsonian matching proposed in Ref.~\refcite{HY:WM} which is done by
matching the axial-vector and vector current correlators derived
from the HLS with those by the operator product expansion (OPE) in
QCD at the matching scale $\Lambda$.
In this section, we present the matching conditions to determine
the bare pion velocity including the effect of Lorentz symmetry
breaking at the bare level following Ref.~\refcite{HKRS:pv}.

In the EFT sector,
pion couples to the longitudinal part of the axial-vector current
correlator $G_A^L$.
We regard $G_A^{L,T}$ as functions of 
$-q^2$ and $| \vec{q} |^2$ instead of $q_0$ and $\vec{q}$,
and expand $G_A^{L}$ in
a Taylor series around $\bar{q}=|\vec{q}|=0$ in $\bar{q}^2/(-q^2)$ as
follows:
\begin{eqnarray}
 G_A^L(-q^2,\bar{q}^2)
 = G_A^{L(0)}(-q^2) + G_A^{L(1)}(-q^2)\bar{q}^2 + \cdots.
\end{eqnarray}
Expanding the axial-vector current correlator derived from the HLS
theory $G_A^{{\rm (HLS)}L}$
in terms of $\bar{q}^2/(-q^2)$, we obtain
\begin{eqnarray}
 &&
 G_A^{{\rm (HLS)}L(0)}(-q^2)
  = \frac{F_{\pi,{\rm bare}}^t F_{\pi,{\rm bare}}^s}{-q^2}
    {}- 2z_{2,{\rm bare}}^L~,
\label{HLS-L(0)}
\\
 &&
 G_A^{{\rm (HLS)}L(1)}(-q^2)
  = \frac{F_{\pi,{\rm bare}}^t F_{\pi,{\rm bare}}^s
    (1 - V_{\pi,{\rm bare}}^2)}{(-q^2)^2},
\label{HLS-L(1)}
\end{eqnarray}
where $z_{2,{\rm bare}}^L$ is the parameter of the higher order term.

On the other hand,
the axial-vector current correlator obtained from the OPE
is given by~\cite{hkl,LM,FLK}
\begin{eqnarray}
 &&
 G_A^{\mu\nu}(q_0,\bar{q})
 = (q^\mu q^\nu - g^{\mu\nu}q^2)\frac{-1}{4}
   \Biggl[ \frac{1}{2\pi^2}
   \Biggl( 1+\frac{\alpha_s}{\pi} \Biggr) \ln \Biggl( \frac{Q^2}{\mu^2}
   \Biggr) + \frac{1}{6Q^4} \Big\langle \frac{\alpha_s}{\pi} G^2
   \Big\rangle_T \nonumber\\
 &&\qquad\qquad{}- \frac{2\pi\alpha_s}{Q^6} \Big\langle
   \Bigl( \bar{u}\gamma_\mu \gamma_5 \lambda^a u -
   \bar{d}\gamma_\mu \gamma_5 \lambda^a d \Bigr)^2 \Big\rangle_T
\nonumber\\
 &&\qquad\qquad{}- \frac{4\pi\alpha_s}{9Q^6}
   \Big\langle
   \Bigl( \bar{u}\gamma_\mu\lambda^a u + \bar{d}\gamma_\mu\lambda^a d
   \Bigr)\sum_q^{u,d,s}\bar{q}\gamma^\mu\lambda^a q \Big\rangle_T
   \Biggr] \nonumber\\
 &&\qquad{}+
[-g^{\mu\nu}q^{\mu_1}q^{\mu_2} + g^{\mu\mu_1}q^\nu q^{\mu_2}
    + q^\mu q^{\mu_1}g^{\nu\mu_2} + g^{\mu\mu_1}g^{\nu\mu_2}Q^2]
\nonumber\\
&&\qquad\qquad{}\times [\frac{4}{Q^4}A_{\mu_1\mu_2}^{4,2}
+\frac{16}{Q^8}q^{\mu_3}q^{\mu_4}A_{\mu_1\mu_2\mu_3\mu_4}^{6,2}] \
,\label{ope-t} 
\end{eqnarray}
where 
$Q^2 = -q^2$, $\tau = d - s$ denotes the twist, and $s = 2k$
is the number of spin indices of the operator of dimension $d$.
In the above expression,  
we restrict ourselves to contributions from the twist 2
$(\tau = 2)$ operators~\footnote{
 The higher the twist of operators becomes,
 the more these operators are suppressed since the dimensions of such
 operators become higher and the power of $1/Q^2$ appear.
}.
$A_{\mu_1 \cdots \mu_{2k}}^{2k + \tau, \tau}$ is the residual Wilson
coefficient times matrix element of dimension $d$ and twist $\tau$.
The general tensor structure of the matrix element of
$A_{\mu_1 \cdots \mu_{2k}}^{2k + \tau, \tau}$ is given in
Ref.~\refcite{hkl}. 

Now we proceed to estimate the pion velocity by matching to QCD.
We require the following matching conditions at $Q^2 = \Lambda^2$:
\begin{eqnarray}
  &&
   Q^2 \frac{d}{d Q^2}G_A^{{\rm (HLS)}L(0)}(Q^2)
   = Q^2 \frac{d}{d Q^2}G_A^{{\rm (OPE)}L(0)}(Q^2),
\nonumber\\
  &&
   G_A^{{\rm (HLS)}L(1)}(Q^2)
   = G_A^{{\rm (OPE)}L(1)}(Q^2).
\end{eqnarray}
They lead to the conditions on the bare pion decay constants as
 \begin{eqnarray}
 &&
  \frac{F_{\pi,{\rm bare}}^t F_{\pi,{\rm bare}}^s}{\Lambda^2}
 = \frac{1}{8\pi^2}\Biggl[ \Bigl( 1 + \frac{\alpha_s}{\pi} \Bigr) +
   \frac{2\pi^2}{3}\frac{\big\langle \frac{\alpha_s}{\pi}G^2
   \big\rangle_T }{\Lambda^4}
   {}+ \pi^3 \frac{1408}{27}\frac{\alpha_s
    \langle \bar{q}q \rangle_T^2}{\Lambda^6}  \Biggr] 
\nonumber\\
 &&\qquad\qquad\qquad\qquad
  {}+ \frac{\pi^2}{15}\frac{T^4}{\Lambda^4}A_{4,2}^\pi
  {}- \frac{16\pi^4}{21}\frac{T^6}{\Lambda^6}A_{6,4}^\pi
 \,\,\equiv G_0,
\nonumber\\
 &&
  \frac{F_{\pi,{\rm bare}}^t F_{\pi,{\rm bare}}^s
 (1 - V_{\rm bare}^2)}{\Lambda^2}
  = \frac{32}{105}\pi^4\frac{T^6}{\Lambda^6} A_{6,4}^\pi,
\end{eqnarray}
where we use the dilute pion-gas approximation in order to
evaluate the matrix element $\langle {\mathcal O} \rangle_T$~\cite{hkl}
in the low temperature region.
{}From these conditions, we obtain the following matching condition to
determine the deviation of the bare pion velocity from the speed of
light in the low temperature region:
\begin{equation}
 \delta_{\rm bare}\equiv 1 - V_{\pi,{\rm bare}}^2
 = \frac{1}{G_0}
   \frac{32}{105}\pi^4\frac{T^6}{\Lambda^6} A_4^{\pi} .
\label{deviation-rho}
\end{equation}
This implies that the intrinsic
temperature dependence starts from the ${\mathcal O}(T^6)$
contribution. On the other hand, the hadronic thermal correction
to the pion velocity starts from the ${\mathcal O}(T^4)$ 
[see Eq.~(\ref{low T})].
Thus the hadronic thermal effect is dominant in
low temperature region.

\vspace*{-0.2cm}

\section{Pion Velocity near the Critical Temperature}

In this section, we extend the matching condition valid at low
temperature, Eq.~(\ref{deviation-rho}), to near the critical
temperature, and determine the bare pion velocity at $T_c$. 

As is
discussed in Ref.~\refcite{HKRS:pv}, we should in principle evaluate
the matrix elements in terms of QCD variables only in order for
performing the Wilsonian matching, which is as yet unavailable from
model-independent QCD calculations.  Therefore, we make an
estimation by extending the dilute gas approximation adopted in
the QCD sum-rule analysis in the low-temperature region to the
critical temperature with including all the light degrees of
freedom expected in the VM. In the HLS/VM theory, both the
longitudinal and transverse vector mesons become massless at the
critical temperature since the HLS gauge coupling constant $g_L$
vanishes. At the critical point, the longitudinal vector meson
which becomes the NG boson $\sigma$ couples to the vector current
whereas the transverse vector mesons decouple from the theory
because of the vanishing $g_L$. Thus we assume that thermal
fluctuations of the system are dominated near $T_c$ not only by
the pions but also by the longitudinal vector mesons. 
We evaluate the thermal matrix elements of the non-scalar operators
in the OPE, by extending the thermal pion gas approximation employed
in Ref.~\refcite{hkl} to the longitudinal vector mesons that figure
in our approach. This is feasible since at the critical
temperature, we expect the equality $A_4^\rho(T_c) = A_4^\pi(T_c)$
to hold as the massless longitudinal vector meson is the chiral partner 
of the pion in the VM. 
It should be noted that, although we use the dilute gas
approximation, the treatment here is already beyond the
low-temperature approximation because
the contribution from vector meson is negligible in the
low-temperature region. Since we treat the pion as a massless
particle in the present analysis, it is reasonable to take
$A_4^\pi(T) \simeq A_4^\pi(T=0)$. We therefore use
\begin{equation}
 A_4^\rho(T) \simeq A_4^{\pi}(T) \simeq A_4^\pi(T=0)
 \quad \mbox{for}\quad T \simeq T_c.
\label{matrix Tc}
\end{equation}
Therefore from Eq.~(\ref{deviation-rho}), we obtain the deviation
$\delta_{\rm bare}$ as
\begin{equation}
 \delta_{\rm bare} = 1 - V_{\pi,{\rm bare}}^2
 = \frac{1}{G_0}
   \frac{32}{105}\pi^4\frac{T^6}{\Lambda^6}
  \Bigl[ A_4^{\pi}+  A_4^{\rho} \Bigr].
\label{deviation-pi-rho}
\end{equation}
This is the matching condition to be used for determining the
value of the bare pion velocity near the critical temperature.

Let us make a rough estimate of $\delta_{\rm bare}$.
For the range of matching scale
$(\Lambda = 0.8 - 1.1\, \mbox{GeV})$, that of QCD scale
$(\Lambda_{QCD} = 0.30 - 0.45\, \mbox{GeV})$ and critical
temperature $(T_c = 0.15 - 0.20\, \mbox{GeV})$, we get
\begin{equation}
 \delta_{\rm bare}(T_c) = 0.0061 - 0.29\,.
\end{equation}
Thus we obtain the $bare$ pion velocity as
$V_{\pi,{\rm bare}}(T_c) = 0.83 - 0.99\,$.
Finally thanks to the non-renormalization property, 
i.e., $v_\pi (T_c)=V_{\pi, {\rm bare}}(T_c)$ given in Eq.~(\ref{phys=bare}), 
we arrive at the physical pion velocity at chiral
restoration:
\begin{equation}
 v_{\pi}(T_c) = 0.83 - 0.99\,,
\end{equation}
to be close to the speed of light.

\vspace*{-0.2cm}

\section{Summary}

In this talk, we started from the Lorentz non-invariant HLS
Lagrangian at bare level and studied the pion velocity at the
critical temperature based on the VM.
We showed that {\it the pion velocity does not
receive either quantum or hadronic corrections 
in the limit $T \to T_c\,$},
which is protected by the VM.
This non-renormalization property
means that it suffices to compute the pion velocity at the level
of bare HLS Lagrangian at the matching scale to arrive at the
$physical$ pion velocity at the critical temperature of chiral 
symmetry restoration.
We derived the matching condition on the $bare$ pion velocity
and finally we found that the pion velocity near $T_c$ is close to
the speed of light, $v_\pi (T) = 0.83 - 0.99\,.$

This is in contrast to the result obtaied from the chiral theory
~\cite{SS},
where the relevent degree of freedom near $T_c$ is only the pion.
Their result is that the pion velocity becomes zero for $T \to T_c$.
Therefore from the experimental data,
we may be able to distinguish which picture is correct, 
$v_\pi \sim 1$ or $v_\pi \to 0$.

\vspace*{-0.2cm}

\section*{Acknowledgment}

The author would like to thank
Professor Masayasu Harada, Doctor Youngman Kim and
Professor Mannque Rho
for many useful discussions and comments.
This work is supported in part by the 21st Century COE
Program of Nagoya University provided by Japan Society for the
Promotion of Science (15COEG01).

\end{document}